\documentclass[aps,pre,twocolumn,groupedaddress,showpacs]{revtex4}

\usepackage{bm,graphicx}
\begin{document}
%

%Title of paper
%
\title{Formation of regular structures in the process of phase separation}
\author{Alexei Krekhov}
%
%\email[]{alexei.krekhov@uni-bayreuth.de}
%
\affiliation{Physikalisches Institut, Universit\"at Bayreuth,
D-95440 Bayreuth, Germany}
\date{ Submitted: July 1, 2008; Revised: December 23, 2008 }
\begin{abstract}
Phase separation under directional quenching has been studied in a 
Cahn-Hilliard model.
In distinct contrast to the disordered patterns which develop under a 
homogeneous quench periodic stripe patterns are generated behind the quench 
front.
Their wavelength is uniquely defined by the velocity of the quench 
interface in a wide range.
Numerical simulations match perfectly analytical results obtained in the 
limit of small and large velocities of the quench interface.
Additional periodic modulation of the quench interface may lead to cellular 
patterns.
The quenching protocols analyzed in this paper are expected to be an 
effective tool in technological applications to design nanostructured 
materials.
\end{abstract}
%

% (PRL) Section: L4 Nonlinear Dynamics, Fluid Dynamics, Classical Optics, 
% etc.
\maketitle
%

% Introduction
%
The dynamics of phase separation in multiphase systems has been 
investigated intensively during the last decades 
\cite{Gunton:1983, Bray:1994}.
A main paradigm are the spontaneously arising spatial concentration 
variations of a characteristic average domain size when a homogeneous 
binary mixture is quenched into the thermodynamically unstable region.
With the progress of time coarsening takes place, i.e., the average domain 
size increases continuously.
The order parameter, usually the relative concentration of the two species, 
obeys a conservation law in contrast to the large number of hydrodynamic 
pattern forming systems \cite{Cross:1993}.
For technological applications it is highly desirable to run a phase 
separation process in a controlled manner to create regular structures.
This is important to design nanostructured materials and 
nanodevices in diverse fields, ranging from bioactive patterns 
\cite{Voros:2005} to polymer electronics \cite{Sirringhaus:2005}.
The arrangement and the size of the domains which form the functional 
elements have a crucial impact on the device performance, e.g., in 
photovoltaics, LEDs, and electronic circuits made of polymer blends 
\cite{Sirringhaus:2005, Coffey:2005, Fichet:2004}.
Previous attempts to manipulate the phase separation in binary 
mixtures by various external fields did not lead to a satisfactory control 
of the pattern morphology 
\cite{Tanaka:1995, Emmott:1996, Berthier:2001, Golovin:2001, Voit:2005}.
A first breakthrough has been established recently by introducing a 
persistent spatially periodic temperature modulation in a model of the 
phase separation in binary mixtures \cite{Krekhov:2004}.
In this case stripe patterns with the periodicity slaved to the externally 
imposed one can be stabilized against coarsening above some critical 
modulation amplitude.
In this paper a new effective mechanism to create periodic stripe patterns 
by directional quenching will be presented.
Their wavelength is uniquely selected by the velocity of quench interface.
If in addition a spatially periodic modulation of the quench interface is 
introduced also cellular patterns can be generated.
The appropriate mean field description of phase separation is commonly 
based on the generic Cahn-Hilliard (CH) model \cite{Cahn:1958} 
(model $B$ \cite{Hohenberg:1977}).
It has been widely used to study the dynamics of phase separation processes 
in a large variety of systems, such as binary alloys, fluid mixtures, and 
polymer blends \cite{Gunton:1983, Bray:1994}.
In the one-dimensional case the CH model is described by the following PDE
\begin{eqnarray}
\label{CHE}
\partial_t u = \partial_{xx} ( -\epsilon u + u^3 -\partial_{xx}u ) \;,
\end{eqnarray}
where $u(x,t)$ is a real order parameter, e.g., in a binary mixture the 
difference of concentration of one species from that at the critical point 
and $\epsilon$ is the control parameter.
According to Eq.~(\ref{CHE}) the spatial average, $\langle u \rangle$,
of $u(x,t)$ is conserved.
To keep the analysis most transparent it is sufficient to concentrate on 
the case of $\langle u \rangle = 0$ (the so-called critical quench).
The homogeneous solution $u=0$ becomes unstable for $\epsilon > 0$ against
linear perturbations $\sim e^{\sigma t + i q x}$ with wavenumber 
$q \in (0, \sqrt{\epsilon})$ and growth rate $\sigma = q^2(\epsilon - q^2)$.
The most unstable (fastest growing) mode is characterized by
$q_m = \sqrt{\epsilon/2}$ with $\sigma_m = \epsilon^2/4$.
For $\epsilon > 0$ a one-parameter family of stationary periodic solutions 
$u_p(x,k)$ of Eq.~(\ref{CHE}), the so-called soliton-lattice solutions, 
exists which can be expressed in terms of the Jacobian elliptic function 
$\text{sn}$ as:
\begin{eqnarray}
\label{SLS}
u_p(x,k) = \frac{\sqrt{2} k}{\xi} \text{sn}(\frac{x}{\xi}, k) \; , \;\;
\xi = \sqrt{\frac{1+k^2}{\epsilon}} \;.
\end{eqnarray}
The modulus $k \in (0, 1)$ is related to the wavenumber $q$:
\begin{eqnarray}
q = \frac{\pi}{2 \text{K}(k) \xi} \; , \;\;
\text{K}(k) = \int\limits_{0}^{\pi/2}
\frac{d \varphi}{\sqrt{1 - k^2 \sin^2 \varphi}} \; ,
\end{eqnarray}
where $\text{K}$ is the complete Jacobian elliptic integral of the
first kind.
It is known, however, that any periodic solution of Eq.~(\ref{CHE}) is 
unstable against period doubling, i.e., against coarsening 
\cite{Langer:1971}.
In the limit $q \ll q_m$ the growth rate of the corresponding destabilizing 
mode is given as \cite{Krekhov:2004, Langer:1971}:
\begin{eqnarray}
\label{eq:sigma_p}
\sigma_p = 
\epsilon^{2} 16 \exp(-2\pi q_m/q)/(\pi q_m/q) \;.
\end{eqnarray}
Note that $\epsilon$ can be scaled out in Eq.~(\ref{CHE}) which is 
reflected in the $\epsilon$-dependence of $u_p$, $q$, and $\sigma_p$.
Coarsening becomes extremely slow and the solutions of type 
(\ref{SLS}) persist for a long time $\Delta t_p \approx 1/\sigma_p$ at 
small $q$ ($k \to 1$).
This situation is favorable to the general goal of controlling
phase separation to achieve long lived periodic structures.
However, with just a homogeneous quench from negative to positive 
values $\epsilon$ in Eq.~(\ref{CHE}) periodic structures with a single $q$ 
can never be obtained:
Any random perturbations about $u=0$ will develop initially into a 
superposition of the fastest growing Fourier modes with average wavenumber 
$\langle q \rangle \approx q_m$.
In the next stage coarsening sets in; $\langle q \rangle$ decreases 
continuously in time and follows a scaling law 
$\langle q \rangle \sim 1/\log t$ \cite{Langer:1971, Kawakatsu:1985}.
Consequently, the question arises, how to control the quenching process 
such that a quasi-stationary periodic solution of type (\ref{SLS}) can be 
generated.
It is well-known that temporal and/or spatial modulations of the 
control parameter serve as a powerful tool to influence the pattern selection 
processes (see, e.g., \cite{vanSaarloos:2003, Rudiger:2007}).
One of the simplest cases is directional quenching where a jump of the control 
parameter is introduced and the arising interface is dragged with a constant 
velocity.
This has motivated us to analyze systematically the impact of the 
directional quenching in the generic CH model.
Thus we have achieved a clear understanding of controlling the generation of 
periodic structures.
In fact, directional quenching has been used in earlier numerical simulations 
of certain phase separation model and some kind of regular patterns have 
been observed \cite{Furukawa:1992, Liu:2000}, to which we will return below.
In the CH model (\ref{CHE}) directional quenching is realized by 
changing $\epsilon$ from a negative value at $x<x_q$ to positive one for 
$x>x_q$, i.e., dividing the system into a stable and unstable region.
The quench interface (referred to as QI in the sequel) at 
$x_q$ is moving in the laboratory frame with a velocity $v$, i.e.,
\begin{eqnarray}
\label{step}
\epsilon(x,t) = \left\{
\begin{array}{rcl}
-\epsilon & , & x < - v t \; , \\
+\epsilon & , & x > - v t \; .
\end{array}
\right.
\end{eqnarray}
%

\begin{figure}[ht]
\includegraphics[width=6.0cm]{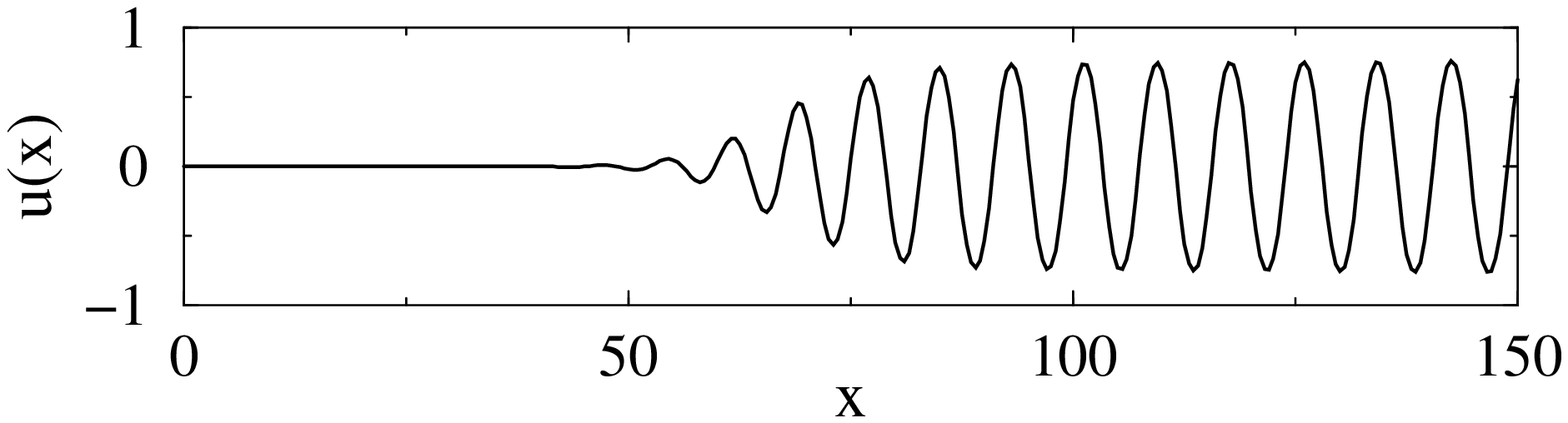} (a)
\includegraphics[width=6.0cm]{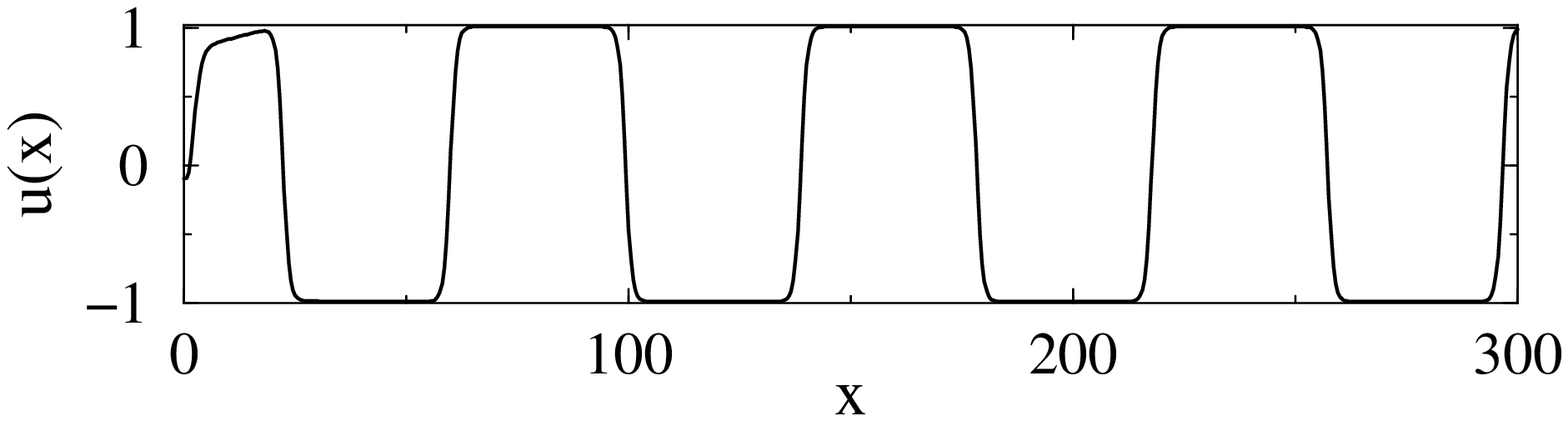} (b)
\caption{Solutions of Eq.~(\protect{\ref{CHE}}) with the QI 
(\protect{\ref{step}}) for $\epsilon=1$ in the comoving frame with the QI 
at $x=0$: 
$v = 2 > v^\star$ (a) and $v=0.02 \ll v^\star$ (b).
Only a part of the system of total length $l_x=4096$ is shown.}
\label{fig:1}
\end{figure}

\begin{figure}[ht]
\includegraphics[width=6.0cm]{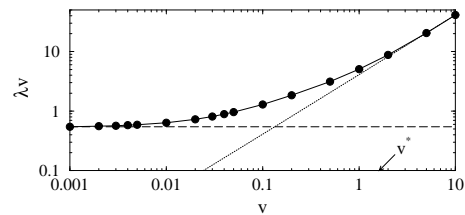}
\caption{Kink separation length $\lambda=\pi/q$ multiplied with the 
velocity $v$ of the QI as a function of $v$ for $\epsilon=1$ (full 
circles with a solid line as a guide to the eye); $v^\star=1.622$ from 
Eq.~(\protect{\ref{v_c_q_c}}). The doted and the dashed lines correspond to 
Eq.~(\protect{\ref{v_c_q_c}}) and Eq.~(\protect{\ref{eq:L}}), respectively.}
\label{fig:2}
\end{figure}

\begin{figure}[ht]
\includegraphics[width=6.0cm]{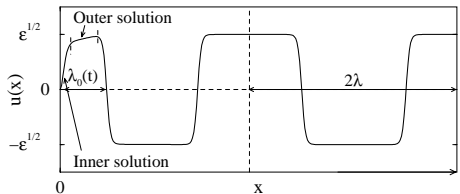}
\caption{Sketch of the solution of Eq.~(\protect{\ref{CHE}}) with the 
QI (\protect{\ref{step}}) in the comoving frame for $x > x_q$ during the slow 
stage at the time shortly before the new kink will be generated at $x_q=0$. 
The domain of $u_{out}(x)$ connecting to $u_{in}$ at $x=x_0$ is marked by the vertical dashes.}
\label{fig:3}
\end{figure}

% 1d case
%
Numerical simulations of the 1d CH model (\ref{CHE}) with the directional
quenching (\ref{step}) demonstrate that a periodic solution develops
behind the QI in the unstable region.
Typical examples for large and small velocities $v$ of the QI
are shown in Fig.~\ref{fig:1}: For $v$ above some critical value, 
$v^\star$, the periodic solutions detaches from the moving QI and the 
wavelength of the solution becomes independent of $v$ 
[Fig.~\ref{fig:1}(a)].
In contrast for $v < v^\star$ the solution remains attached to the QI with 
the wavelength uniquely determined by $v$.
Decreasing $v$ the solution develops into a periodic kink lattice (sharp 
changes between $u=\pm \sqrt{\epsilon}$) where new kinks are continuously 
generated at $x=x_q(t)=-v t$ [Fig.~\ref{fig:1}(b)].
The period of the solution, $2\lambda$, turns out to be uniquely defined by 
velocity of the QI and is shown in Fig.~\ref{fig:2}.
For $v \to 0$ one has $\lambda \sim 1/v$ whereas for $v > v^\star$ one 
finds $\lambda = \pi/q^\star$.
Although the periodic solutions far away from the moving QI 
are in principle unstable against period doubling, the coarsening is 
extremely slow for patterns generated with $q \ll q_m$ [see 
Eq.~(\ref{eq:sigma_p})].
Thus the extension $L_p$ of the (quasi-ideal) periodic solution behind the 
QI can be estimated as 
$L_p = v \Delta t_p \approx v/\sigma_p$ where $\sigma_p$ is the growth rate 
of the unstable period doubling mode given in Eq.~(\ref{eq:sigma_p}).
The two limiting cases of large and small velocity $v$ of the QI can be 
captured analytically.
For large $v$ we consider for instance the initial condition $u=0$ 
everywhere except a hump $u>0$ localized near $x=0$.
Then the time evolution of this initial perturbation is governed by the 
motion of wave fronts  to the left and to the right with a well-defined 
velocity $v^\star$ and wavenumber $q^\star$.
These quantities can be calculated by a linear stability analysis of the 
leading edge of the front in the comoving frame.
One arrives thus at the so-called marginal stability criteria
\cite{Ben-Jacob:1985, vanSaarloos:2003}:
\begin{eqnarray}
\label{MST}
\frac{d \tilde{\sigma}(q_0, v^\star)}{d q} = 0 , \;
\text{Re}[\tilde{\sigma}(q_0, v^\star)] = 0 , \;
q^\star = \frac{\text{Im}[\tilde{\sigma}(q_0, v^\star)]}{v^\star} ,
\end{eqnarray}
where $\tilde{\sigma}(q,v) = q^2(\epsilon-q^2) + i q v$ and $q_0$ is a 
complex number (saddle point).
Eqs.~(\ref{MST}) lead to the solution
\begin{eqnarray}
\label{v_c_q_c}
&&v^\star = \frac{\sqrt{7}+2}{3}
\left( \frac{2}{3} (\sqrt{7}-1) \right)^{1/2} \epsilon^{3/2}
\; ,
\nonumber \\
&&q^\star =
\frac{3(\sqrt{7}+3)^{3/2}}{8\sqrt{2}(\sqrt{7}+2)} \epsilon^{1/2}
\; .
\end{eqnarray}
The phase velocity and the wavenumber of the propagating periodic solutions 
obtained from the numerical simulations of Eq.~(\ref{CHE}) for 
$v > v^\star$, which do not depend on $v$, agree perfectly with $v^\star$ 
and $q^\star$ given by Eq.~(\ref{v_c_q_c}) (Fig.~\ref{fig:2}).
In the opposite limit $v \to 0$ our starting point is a particular 
stationary solution of Eq.~(\ref{CHE}) for $v = 0$ interpolating between 
$u=0$ at $x < 0$ and $u=\sqrt{\epsilon}$ at $x > 0$ which is characterized 
by a sharp front at $x \approx 0$.
If the QI according to Eq.~(\ref{step}) starts to move, the 
sharp front would initially follow.
But since the spatial average $\langle u \rangle$ is conserved regions 
with $u < 0$ have to be generated in the region $x > x_q$ which leads to
the formation of kink lattice [Fig.~\ref{fig:1}(b)].
Its formation can be understood in terms of a fast switching stage and slow 
pulling stage: 
first a new kink is generated in a short time at $x \approx x_q$.
During the slow stage this kink is pulled by the QI whereby its amplitude 
and the distance to the next kink behind, $\lambda_0(t)$, increase until 
$\lambda_0$ exceeds some limiting value $\lambda_{0, max}$ and then a new 
kink is generated (Fig.~\ref{fig:3}).
Repeating this process a regular kink lattice develops in the wake of the 
QI with a kink separation length $\lambda$ (i.e., with the period 
$2\lambda$), which is uniquely determined by the velocity $v$ of the moving 
QI (Fig.~\ref{fig:2}).
During the slow stage the solution of Eq.~(\ref{CHE}) in the interval 
$0 \le x < \lambda_0(t)$ can be described in a comoving frame 
$x \to x + v t$ by
\begin{eqnarray}
\label{eq:slow}
v u = \partial_x ( -\epsilon u + u^3 -\partial_{xx}u ) \;,
\end{eqnarray}
which is obtained from Eq.~(\ref{CHE}) by an $x$ integration with the 
boundary conditions $u=0$ and zero flux 
$\partial_x ( -\epsilon u + u^3 -\partial_{xx}u )=0$
for $x=x_q=0$.
In addition the explicit time dependence $\partial_t u$ can be safely 
neglected since it is only relevant during the fast switching stage.
The solution of Eq.~(\ref{eq:slow}) can be separated into a strongly 
varying inner solution $u_{in}(x)$ ($0 \le x \le x_0$, first kink) and to 
the almost flat outer solution $u_{out}(x)$ ($x > x_0$) (Fig.~\ref{fig:3}).
For calculating $u_{in}$ the term $v u \approx 0$ in Eq.~(\ref{eq:slow}) and 
for $u_{out}$ the term $\partial_{xx} u \approx 0$ can be 
neglected which leads to
\begin{eqnarray}
\label{eq:inner_outer_sol}
&& (\partial_x u_{in})^2 + u_{in}^2 (\epsilon  - \frac{u_{in}^2}{2}) =
C + 2 u_{in} D \;,
\nonumber\\
&& \; C = (\partial_x u_{in})^2 \bigr|_{x=0} \;, \;\;
D = 
[ \partial_{xx} u_{in} + u_{in} (\epsilon - u_{in}^2) ] \bigr|_{x=x_0} \;,
\nonumber\\
&& \partial_x u_{out} = \frac{v u_{out}}{3 u_{out}^2 - \epsilon} \;.
\end{eqnarray}
The inner and outer solutions are matched at the point 
$x=x_0$ chosen such that ``flatness'' conditions 
$\partial_x u_{in} = \partial_{xx} u_{in} = 0$ hold.
This gives for $u_0=u_{in}(x_0)$ the following expression
\begin{eqnarray}
\label{eq:inner_sol}
u_0^2 = \frac{\epsilon}{3} + \sqrt{ \left( \frac{\epsilon}{3} \right)^2
+ \frac{2}{3}C } \;.
% \;\; C = [\partial_x u \bigr|_{x=0}]^2 > 0 \;.
\end{eqnarray}
Starting from $u=u_0$ the outer solution $u_{out}$ will grow until at 
$x=\lambda_0$ (second kink) the maximal possible amplitude 
$u_{max}=\sqrt{\epsilon}$ is reached.
The maximal interval $\lambda_{0, max}=x_{max}-x_{min}$ where the outer 
solution is supported corresponds obviously to the minimal value 
$u_{min}=\sqrt{2 \epsilon/3}$ of $u_0$ [$C=0$ in Eq.~(\ref{eq:inner_sol})].
Let us now calculate the equilibrium distance between kinks $\lambda$ 
(e.g., the distance between the third and the fourth kink in 
Fig.~\ref{fig:3}).
During the slow pulling stage the distance between the second and the third 
kink increases until it reaches its maximal value $\lambda$ when a new kink 
is generated at the QI.
Inspection of Fig.~\ref{fig:3} shows that then because of 
$\langle u \rangle = 0$ the area $\lambda \sqrt{\epsilon}$ under the curve 
between two neighboring kinks away from the QI equals twice the area 
$S_{max}$ under the curve between the first and the second kink with
\begin{eqnarray}
S_{max} \equiv
\int_{x_{min}}^{x_{max}} u dx =
\int_{u_{min}}^{u_{max}} u \left(\frac{du}{dx}\right)^{-1} du \;.
\end{eqnarray}
With the use of the outer solution (\ref{eq:inner_outer_sol}) one obtains from 
$\lambda \sqrt{\epsilon} = 2 S_{max}$
\begin{eqnarray}
\label{eq:L}
\lambda = 2 \frac{\sqrt{6}}{9} \frac{\epsilon}{v}
= 0.544 \frac{\epsilon}{v}
\end{eqnarray}
in perfect agreement with the results of numerical simulations in the limit
$v \to 0$ (Fig.~\ref{fig:2}).
The generalization of the analysis to the off-critical quench, 
$\langle u \rangle \ne 0$, is straightforward.
The expressions (\ref{v_c_q_c}) for $v^\star$ and $q^\star$ hold 
except the replacement $\epsilon \to \epsilon - 3 \langle u \rangle^2$.
In the limit $v \to 0$ the distance $\lambda_{+}$ between two kinks in the 
region $u>0$, and $\lambda_{-}$ for $u<0$, respectively, become different.
We find 
$\lambda_{+} - \lambda_{-} = \langle u \rangle \Lambda /\sqrt{\epsilon}$
and for the resulting period $\Lambda$ of the kink lattice
\begin{eqnarray}
\label{eq:Lmbd}
\Lambda \equiv \lambda_{+} + \lambda_{-} =
\frac{2}{v} \left[ 2 \frac{\sqrt{6}}{9} \epsilon 
+ (8 - 25\frac{\sqrt{6}}{9}) \langle u \rangle^2 \right] \;.
\end{eqnarray}
As in the case $\langle u \rangle =0$, we found Eq.~(\ref{eq:Lmbd}) confirmed 
in numerical simulations of Eq.~(\ref{CHE}) for different values of $\langle u 
\rangle$ in the limit $v \to 0$.
%

%

\begin{figure}[ht]
\includegraphics[width=3.2cm]{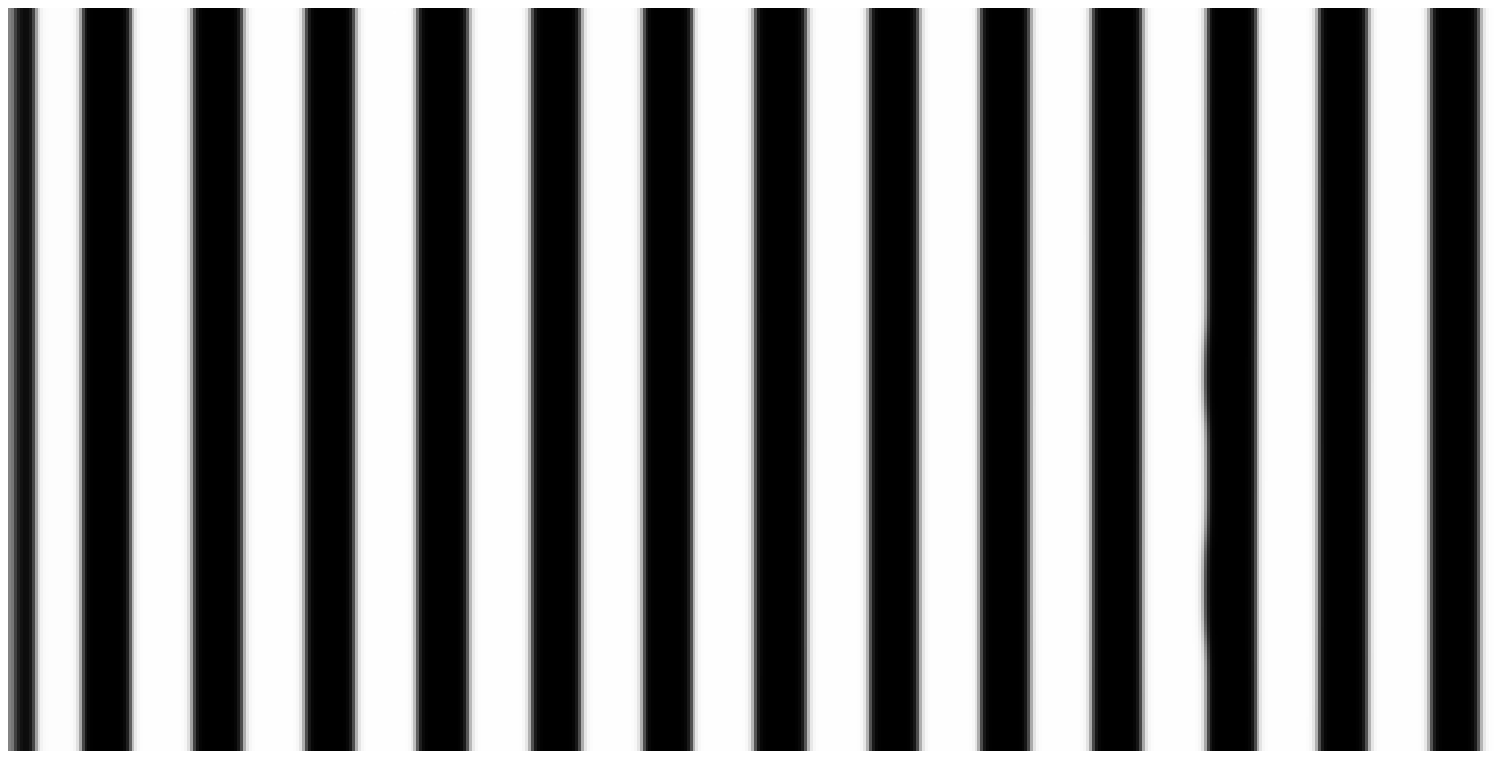} (a)
\includegraphics[width=3.2cm]{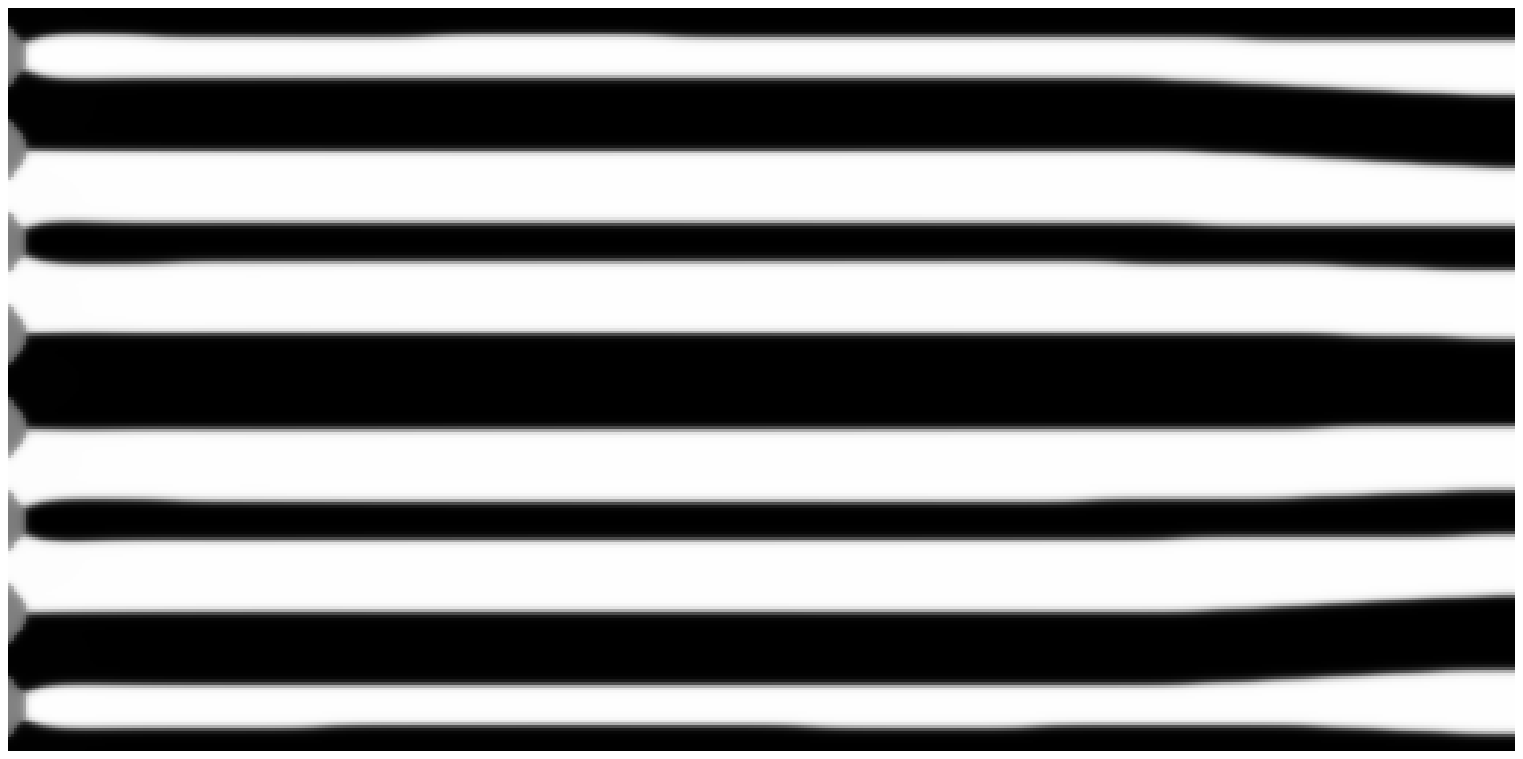} (b)
\includegraphics[width=3.2cm]{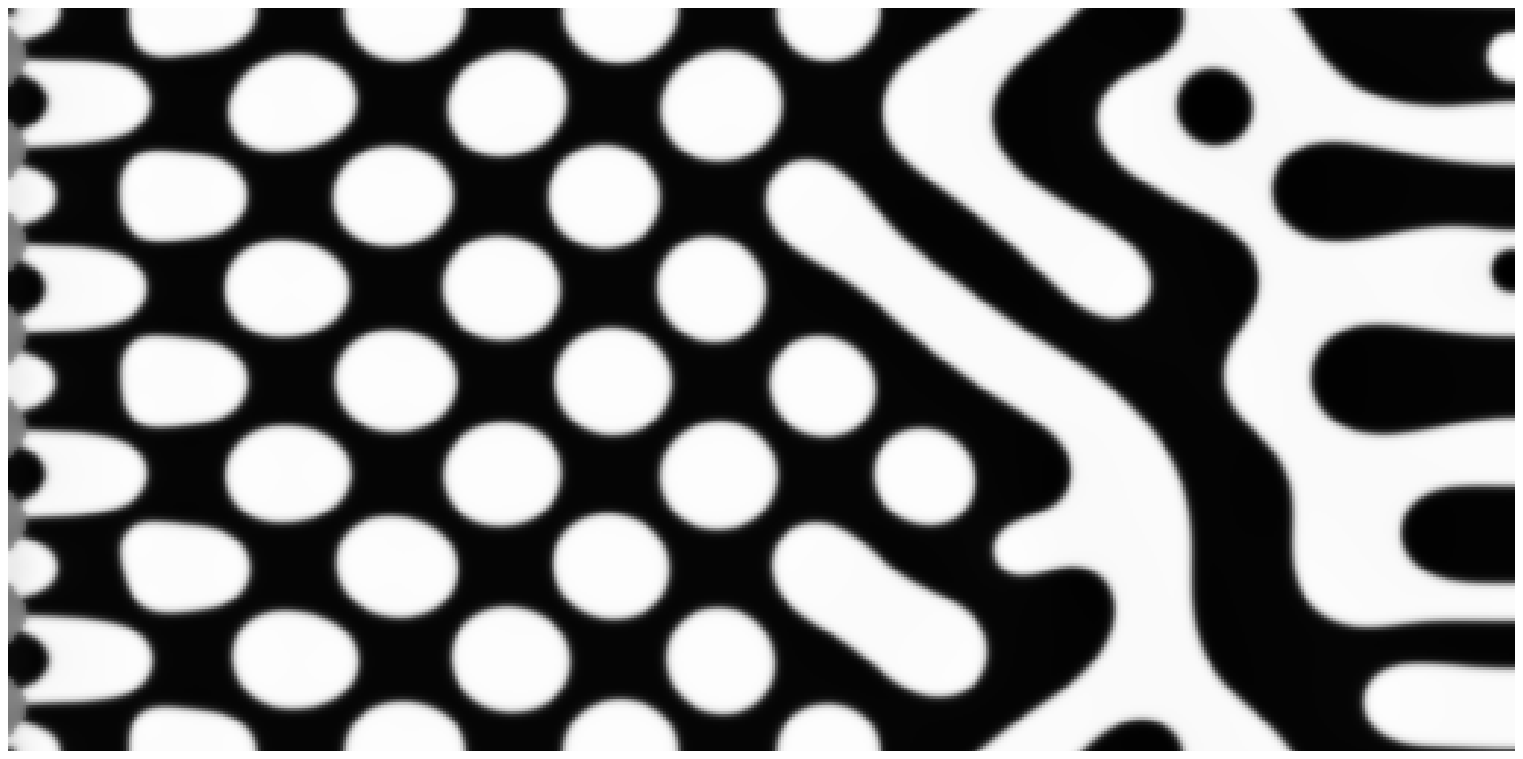} (c)
\includegraphics[width=3.2cm]{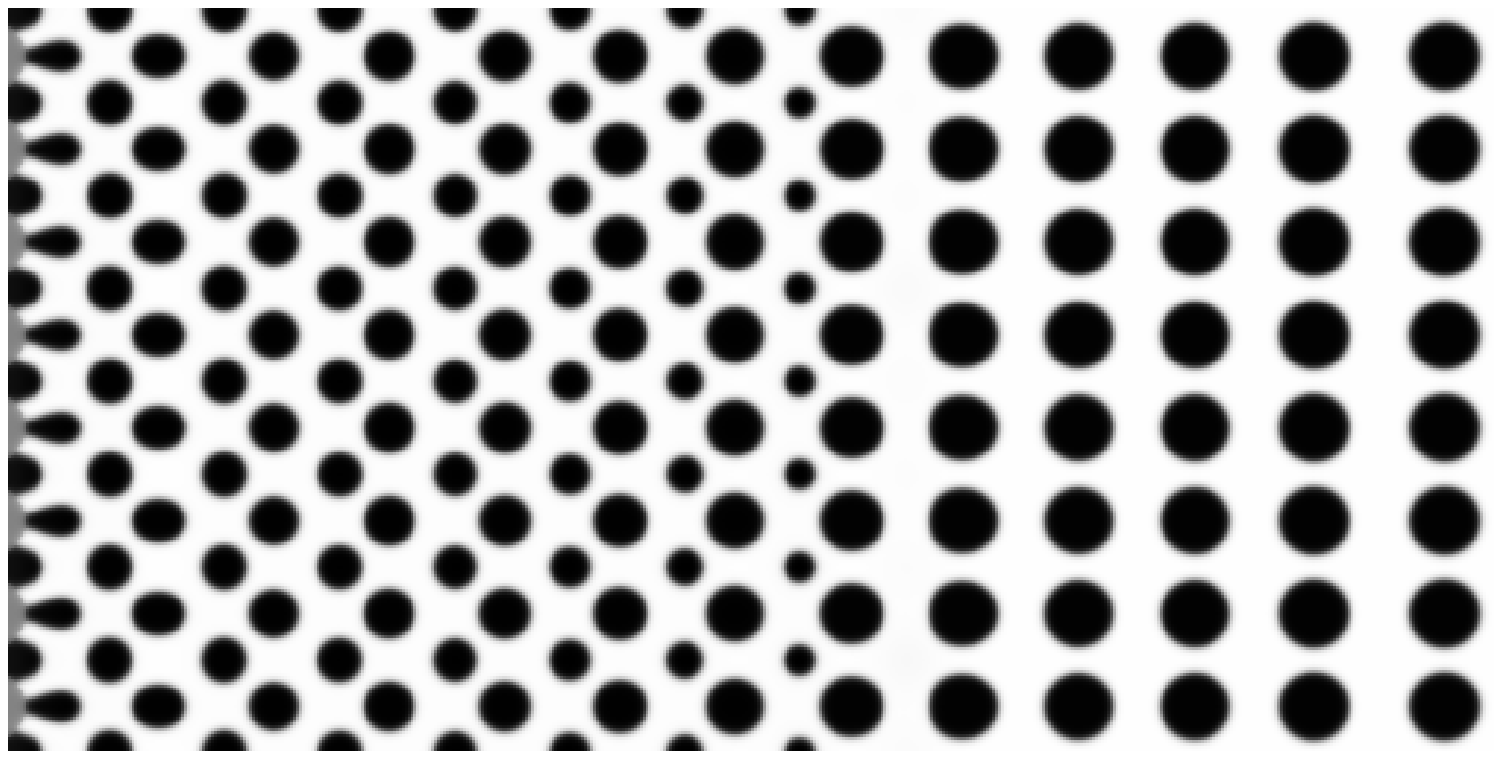} (d)
\caption{Snapshots of the phase separation in 2d at the time when the QI 
almost reaches the left boundary.
Straight QI (\protect{\ref{step}}) for $\epsilon=1$, 
$\langle u \rangle = 0.1$, $v=0.05$ (a).
Modulated QI (\protect{\ref{mod_step}}) for $\epsilon=1$, $a=4$, $p=\pi/16$:
$\langle u \rangle = 0$ and 
$v=0.01$ (b), $v=0.055$ (c);
$\langle u \rangle = 0.1$, $v=0.1$ (d).}
\label{fig:4}
\end{figure}

% 2d case
%
Let us now switch to the two-dimensional case $u(x,y,t)$ where we study 
first numerically the 2d version of CH equation (\ref{CHE}):
\begin{eqnarray}
\label{eq:CH2d}
\partial_t u =
\nabla^2 ( -\epsilon u + u^3  - \nabla^2 u )
\end{eqnarray}
with the moving QI (\ref{step}).
Zero flux boundary conditions have been used at $x=0, l_x$ and periodic 
boundary conditions at $y=0, l_y$.
Initially the QI is located at $x_q=l_x$ moving from right to left.
The system size was $l_x=512$, $l_y=256$ and we start with the homogeneous 
solution $u=\langle u \rangle$ with small superimposed noise of the strength 
$\delta u$ where $\delta u \ll \sqrt{\epsilon}$ and $\delta u \ll \langle u 
\rangle$.
Thus the well-known Ginzburg criterion, necessary for the validity of a 
mean-field description of a phase separation process \cite{Binder:1983}, is 
satisfied: in fact the dynamics does not depend on the particular choice of 
$\delta u$.
For the off-critical quench $\langle u \rangle \ne 0$ when $v < v^\star$ 
always regular stripe patterns with domains parallel to the QI were found 
[see, e.g., Fig.~\ref{fig:4}(a)].
This situation is covered by an 1d analysis presented before where the 
period of the structure is uniquely determined by the velocity of the 
QI.
In the limit $v \to 0$ the period of the patterns found in our numerical 
simulations agree with (\ref{eq:Lmbd}).
For $v > v^\star$ irregular coarsening patterns similar to the case of a 
spatially homogeneous quench have been observed.
In contrast, in the case of the critical quench $\langle u \rangle = 0$, 
the orientation of the domains depends on the velocity of the QI.
At small $v$ periodic patterns with domains perpendicular to the QI are 
formed [similar to Fig.~\ref{fig:4}(b)].
Then for $v$ above $v_c \approx 0.45$ the 1d stripe patterns parallel to 
the QI appear as in the off-critical case described above.
Finally $v > v^\star$ leads eventually to irregular patterns similar to the
case of a spatially homogeneous quench.
The earlier studies of a related model subjected to directional quenching in 
2d \cite{Furukawa:1992, Liu:2000}, which demonstrate the existence of regular 
patterns as well, were purely numerical and do not give real insight into the 
pattern selection mechanisms.
In particular, for the relevant off-critical case a noise strength $\delta u > 
\langle u \rangle$, violating the Ginzburg criterion, was chosen 
\cite{Liu:2000}.
Consequently, for instance, the change of the domain orientation from 
perpendicular to parallel with increasing $v$ reported in \cite{Liu:2000} is 
considered as an artefact.
Finally we have studied the influence of a periodic modulation of the QI which 
reads as follows
\begin{eqnarray}
\label{mod_step}
\epsilon(x,y,t) = \left\{
\begin{array}{rcl}
-\epsilon & , & x < l_x + a \cos(p y) - v t \; , \\
+\epsilon & , & x > l_x + a \cos(p y) - v t \; .
\end{array}
\right.
\end{eqnarray}
In the case of a critical quench we found that the velocity $v_c$ at which 
the transition from parallel stripe patterns [similar to 
Fig.~\ref{fig:4}(a)] to perpendicular ones [Fig.~\ref{fig:4}(b)] occurs, 
depends on the modulation amplitude $a$.
This dependence is very strong for values $a$ of the order of the typical 
domain size at the initial stage of phase separation 
($a \sim \lambda_m=\pi/q_m$).
Furthermore, $v_c$ decreases with decreasing modulation wavenumber $p$.
For $p$ smaller than the wavenumber $q_m$ of the fastest growing mode 
patterns with a cellular morphology forming behind the moving QI have 
been found [Fig.~\ref{fig:4}(c)].
In the case of an off-critical quench we found that
$\langle u \rangle \ne 0$ favors the formation of regular cellular 
planforms [Fig.~\ref{fig:4}(d)] at intermediate QI velocities, 
in analogy to the transition from parallel to perpendicular stripes for 
$\langle u \rangle = 0$.
%
%

% Conclusion
%
As a main result of the paper we have demonstrated that directional 
quenching in CH model leads to the formation of periodic solutions with the 
wavelength uniquely selected by the velocity of quench interface.
This is in contrast to Ginzburg-Landau models with a nonconserved 
order parameter where a moving quench interface leaves no kinks behind in 
the wake \cite{Kopnin:1999}.
The 1d CH model has recently found a new interesting application in liquid 
crystals to describe Ising walls between symmetry degenerated director 
configurations \cite{Chevallard:2002, Nagaya:2004}.
It should be possible to verify our predictions on the wavelength selection 
by dragging the liquid crystal layer into a homogeneous magnetic field.
In conclusion, controlling phase separation by directional quenching turns 
out to be a promising tool to create regular structures in material science.
Although slow coarsening cannot be avoided by directional quenching in 
principle, long lived periodic patterns can be ``frozen in'', e.g., by 
a deep quench, induced polymerization, chemical treatment etc.
In this respect it would be certainly rewarding to study in more depth the 
2d case where interesting cellular structures have been detected and in 
particular the 3d case.
I am deeply indebted to the late Lorenz Kramer from whom I have benefited a 
lot regarding the concept used in this paper.
It is a pleasure to thank W. Pesch and W. Zimmermann for stimulating 
discussions and for critically reading the manuscript.
Financial support by Deutsche Forschungsgemeinschaft Grant SFB~481 is
gratefully acknowledged.
%

%
% References
%

%

\end{document}